\documentclass{icrc29}
\usepackage{graphicx,amssymb,amsmath,times}
\begin{document}
\title[Observations of Local Group Galaxies with the Whipple 10m Gamma-Ray
  Telescope]{Search for TeV Radiation from Selected Local Group Galaxies}
\author[J. Hall et al.] {J. Hall$^a$ for the VERITAS Collaboration \\
        (a) University of Utah Department of Physics, University of Utah, 
            Salt Lake City, Utah 84112 USA\\
}
\presenter{Presenter: J. Boyle (jeter@physics.utah.edu), \  
ind-student-D-abs1-og21-poster}
\maketitle
\begin{abstract}
Some candidate dark matter particles, such as neutralinos in supersymmetry,
would annihilate producing GeV/TeV gamma rays. We report on recent
observations of two dwarf spheroid galaxies, Draco, Ursa Minor, 
the compact elliptical galaxy M32,
and the spiral galaxy M33 with the Whipple 10m
gamma-ray telescope. No significant signal was found, and we derived
upper limits for the gamma-ray flux from each object.  We discuss our
astrophysical selection criteria for these galaxies in the context of an
indirect search for dark matter and the implications of these flux upper
limits on the density of neutralinos.
\end{abstract}
\section{Introduction}
One candidate for the dark matter (DM) motivated by particle physics is the
lightest neutral supersymmetric particle, the neutralino.
Supersymmetry is theoretically favored because it naturally leads to
a unification of energy scales for all the fundamental 
forces and is a natural symmetry for grand unification
theories such as string theory.  The
neutralino is a majorana particle, so it will annihilate and create
particles with energies similar to the mass of the 
neutralino.  Some
of these particles would be photons, charged particles that would start
emitting and scattering light, and neutral pions that would
decay into photons illuminating a neutralino star at 
energies near the neutralino mass.  This radiation could be detectable
by ground and/or orbiting gamma ray observatories\cite{bergstrom98}.
\par
The Galactic Center (GC) was proposed early as a source of gamma-ray radiation
from annihilation of DM\cite{silk87}, but recent TeV observations of the GC 
has shown a point source coincident with the GC
\cite{kosack04,tsuchiya04,aha04} which appears to be a background for any
observations of an annihilation signal.  The next candidates for
these observations are the local group galaxies \cite{baltz00}.  We carried
out an observational campaign on selected local group galaxies with the Whipple
10m gamma-ray telescope on Mt. Hopkins in Arizona, USA.  No significant signal
was found and we derive upper limits for the gamma-ray flux from these
objects.  The poster will include a spectrally dependent analysis and an
extended source analysis.  
\section{Source Selection}
The expected flux from neutralino annihiliation is
\begin{equation}
\frac{d\Phi}{dE\,d\Omega} = \frac{dN}{dE} \, 
\frac{\langle\sigma v\rangle}{8\pi m^2_\chi} \, \frac{dJ(\theta,\phi)}{d\Omega}
\end{equation}
with 
the cross-section $\langle \sigma v \rangle$, the particle mass m, and the
spectrum produced $dN / dE$.  The source profile is
related to the density profile through $dJ / d\Omega = \int dl \ \rho^2$
where integral is taken over the line of sight.
This flux factorization shows the terms depending on the particle physics,
separated from the astrophysical considerations which are contained in 
$dJ/d\Omega$.  For a source small compared to the angular
resolution of the detector at a distance D, the integrated J can be expressed
as an integral
over the volume of the source as
\begin{equation}
J \equiv \int\limits_{\rm source} d\Omega \, \frac{dJ(\theta,\phi)}{d\Omega}
= \frac{1}{D^2} \int\limits_{\rm source} dV \rho^2.
\label{J}
\end{equation}
This integrated $J$ is a parameter that embodies the two astrophysical
considerations for DM annihilation flux, the distance to the source
and the density profile of the DM particles.  Close, dense DM
clumps are therefore the best targets for
observation.  We choose the local group galaxies in order to compete with the
$D^{-2}$ factor in $J$.  Additionally we tried to choose a variety of galaxy
morphologies, because the baryonic matter in the galaxy can disrupt any DM
cusp through heating. Black holes are also expected to have a large
effect on the inner regions of the DM profile.  Adiabatic growth of a black
hole can lead to a spike in the DM profile\cite{gondolo99}.  However any large
merging event between two supermassive black holes should wash out any cusp in
the profile. 
\par
We focused this search on local group galaxies.  The Galactic Center gamma-ray
flux is probably not due to DM annihilation and so represents a background for
a search for an annihilation signal.  Additionally annihilation in the halo of
the Milky Way could be large compared to the field of view ($\sim$3-5$^\circ$)
and ground based imaging instruments have less sensitivity for such sources. 
\subsection{Draco and Ursa Minor Dwarf Galaxies}
Dark matter rich Milky Way satellite Dwarf Galaxies are favored due to their
proximity, high mass to light ratio, and the possibility of self-interacting
DM.  If the DM is non-interacting the cores of these galaxies will not evolve
within the Hubble time.  The density of these smaller halos is found to be
higher in simulations due to their earlier production epochs.  The dwarf
galaxies we chose were smaller than the field of view (2.4$^\circ$).  In
the visible wavelengths they were each $\sim$0.5$^\circ$ in diameter.
\par
The distance to Draco is measured to be 76 $\pm$ 6 kpc in \cite{bonanos04}
using 
variable stars. In \cite{kleyna02} they measure the radial velocity
dispersion of giant stars in Draco.  They find the rotation curve is flat or
slightly increasing away from the core.  The best fit DM halo profile is a
nearly isotropic sphere (r $\sim$ r$^{-0.13}$) of DM with a total mass of
$8^{+3}_{-2} \times 10^7$ M$_\odot$ in the inner 0.5$^\circ$.
\par
The distance to Ursa Minor is 69 $\pm$ 4 kpc \cite{mighell99}.
Ursa Minor has clear substructure observed in the inner $\sim$
10' \cite{kleyna98}.  The
existence of this structure and an associated second population of stars is a
mystery as this structure should only exist for a few hundred million years
based on observations of stellar proper motions\cite{eskridge01}. This raises
the possibility of self-interacting DM. A comparison of the observed
astrophysical properties of Draco and Ursa Minor is given in
\cite{bellazzini02}.
\subsection{M31-M33}
The first rotation curve was made of M31 which was one of the earliest
evidence of DM on galactic scales\cite{babcock}.  Andromeda is a galaxy larger
than the Milky Way at a distance of 785 $\pm$ 25\cite{mcconnachie05}.
Unfortunately there is
evidence that M31 is undergoing a merger with two distinct populations of
stars and a double nucleus.  This merging process should wash out any cusp in
the DM density profile so M31 was not chosen for this survey.
\par
M32 is the closest compact elliptical galaxy at a distance similar to 
M31.  Stellar velocities as well as
gas dynamics measure a single supermassive compact object, $\sim$3.6 x 10$^6$
M$_\odot$, in the core of M32 \cite{joseph01}. Hubble observations of the core
of M32 suggest a density $>$2 x 10$^6$ M$_\odot$ \cite{lauer98}.  
The core of M32 is relaxed and so it is a good candidate for structures such
as DM spikes around the central black hole.  M32 extends about 10' across the
sky so any emission should be point-like.
\par
M33 is a spiral galaxy at a distance of 809 $\pm$ 24 kpc measured using 
red giant stars \cite{mcconnachie05}.  The mass and density profile of
the DM halo of M33 derived from rotation curves is well studied due to its
proximity and orientation\cite{corbelli00}.  M33 does not have a large black
hole in the center \cite{gebhardt01} which could lead to larger densities of DM
if supermassive black hole formation disrupts the halo.  M33 has an angular
size of ~1$^\circ$ so it is contained in the field of view, but it may be
an extended source.  However most halos that would be visible with current
sensitivities would be point-like.
\begin{table}
\begin{center}
\label{T1}
\begin{tabular}{l|c|c|cc|cc}
\hline
\hline
Source&RA&Dec&ON/TRK (hrs)&OFF (hrs)&ON used (hrs)&OFF used (hrs)\\
\hline
Draco&17 20 14&+57 55&14.5&7.5&10.3&5.6\\
Ursa Minor&15 09 10&+67 13&18.4&8.4&7.0&7.0\\
M32&00 42 00&+40 52&10.3&9.3&8.9&8.9\\
M33& 01 33 51&+30 39&18.6&9.1&8.7&8.7\\
\hline
\end{tabular}
\caption{The total exposure and the exposure used for this analysis 
on Draco, Ursa Minor, M32, and M33.  The data were rejected most often
for weather problems and high voltage failure.  The data for Ursa Minor
and Draco were taken in the 2002 observing season.  The M33 dataset was
taken in the 2004 observing season.  The data for M32 was taken during both of
these seasons.}
\end{center}
\end{table}
\section{Analysis}
The data were taken on clear, moonless periods of the night during two
observing seasons.  The data on Ursa Minor and Draco
were taken in 2002-2003 and the data on M32 were taken during 2004-2005.  The
M33 exposure was split evenly between these two periods.  The Whipple 10m
gamma-ray telescope records 40 ns exposures of extended air showers, and
through the analysis electromagnetic showers are chosen and the anisotropy of
these showers is studied for any significant excesses.
\par
For this paper we use the traditional analysis method for the Whipple 10m
telescope.  The first step involves data selection for quality by studying the
observing log of the telescope for any data problems such as clouds or 
high voltage failure.  Once the data is accepted as high quality the pixels
are padded to make up for any differences in field brightness.  Then the
images are cleaned to remove the signal from random fluctuations.  Finally we
use the standard cuts to select the electromagnetic showers from the
hadronic background and the random triggers.
\par
For the hypothesis of a point source in the center of the field of view we
make a cut on the directions of the air showers to within $\sim0.3^\circ$. 
Contemporaneous data on the Crab supernova remnant yields a signal with a
significance of 5.5 $\sigma/\sqrt{\rm hour}$
at a rate of $2.1 \pm 0.1$ photons$/$min with an energy threshold of $\sim400$
GeV. 
A more complete description of the Whipple standard operations are given by
\cite{moh98}. 
\par
The results from the point source analysis are shown in table
\ref{ResultsTable}.  These results will be used
to place an upper limit on the density of neutralinos, specifically $J$
defined in equation \ref{J}, given a particle model of $\langle \sigma v
\rangle$, $m$, and $dN / dE$.  Limits on $J$ will be presented in the poster. 
\section{Conclusions}
We have reported on a campaign to observe DM annihilation at TeV local group
galaxies with the Whipple 10m telescope.  Results from the standard point
source analysis have been presented.  No significant radiation was detected
and upper limits were derived.  A spectrally dependent analysis as well as an
extended source analysis will be presented in the poster.  A campaign to
observe possible DM annihilation is ongoing within the VERITAS collaboration
and observations are being expanded to cover systems such as clusters.
Additionally VERITAS will have a better sensitivity and lower energy threshold
than the Whipple 10m telescope which should allow stronger constraints on the
density of neutralinos. 
\section{Acknowledgments}
We thank the VERITAS collaboration and the University of Utah for help in
operation and analysis of the Whipple 10m telescope. This research is
supported by the National Science Foundation under NSF Grant \#0079704, and by
funding from the U.S. Department of Energy,  the Smithsonian Institution, 
NSERC in Canada, PPARC in the UK, and Science Foundation Ireland.

\begin{table}
\begin{center}
\begin{tabular}{c|c|c|c}
\hline
Object & Significance [$\sigma$] & Upper Limit [$\gamma$ min$^{-1}$] &
Upper Limit [ Crab Flux ] \\
\hline
Draco       & -2.02 & 0.07 &  0.03\\
Ursa Minor  & +0.79 & 0.24 &  0.12\\
M32         & -0.42 & 0.20 &  0.10\\
M33         & +1.23 & 0.25 &  0.12\\
\hline

\hline
\end{tabular}
\caption{The 95\% confidence level upper limits for the point source analysis
  of the data using the standard Whipple 10m analysis. The Crab flux
  normalization is determined by passing contemporaneous Crab data events
  through the same cuts. The energy range of these cuts will be described in
  the poster.}
\label{ResultsTable}
\end{center}
\end{table}

\end{document}